\pdfoutput=1
\documentclass{sigchi-ext}
\usepackage[T1]{fontenc}
\usepackage{textcomp}
\usepackage[scaled=.92]{helvet} 
\usepackage{graphicx} 
\usepackage{balance}  
\usepackage{booktabs} 
\usepackage{ccicons}  
\usepackage{ragged2e} 
\usepackage{subcaption}

\newcommand{\dquote}[1]{``#1''}

\title{Tangible Holograms:\\ Towards Mobile Physical Augmentation of Virtual Objects}

\numberofauthors{6}
\author{%
  \alignauthor{%
    \textbf{Beat Signer}\\
    \affaddr{WISE Lab} \\
		\affaddr{Vrije Universiteit Brussel} \\
    \affaddr{Pleinlaan 2, 1050 Brussels} \\
		\affaddr{Belgium} \\
    \affaddr{bsigner@vub.ac.be} }\alignauthor{%
    \textbf{Timothy J. Curtin}\\
    \affaddr{WISE Lab} \\
		\affaddr{Vrije Universiteit Brussel} \\
    \affaddr{Pleinlaan 2, 1050 Brussels} \\
		\affaddr{Belgium} \\
    \email{tcurtin@vub.ac.be} } \vfil \alignauthor{%
 } }

\def\plaintitle{Tangible Holograms: Towards Mobile Physical Augmentation of Virtual Objects} \def\plainauthor{Beat Signer, Timothy J. Curtin}
\def\plainkeywords{Mixed reality; tangible holograms; data physicalisation; \mbox{programmable} matter; shape-changing interfaces.}

\hypersetup{%
  pdftitle={\plaintitle}, pdfauthor={\plainauthor},
  pdfkeywords={\plainkeywords}, }

\begin{document}

\maketitle

\RaggedRight{} 


\begin{abstract}
The last two decades have seen the emergence and steady development of tangible user interfaces. While most of these interfaces are applied for input---with output still on traditional computer screens---the goal of programmable matter and actuated shape-changing materials is to directly use the physical objects for visual or tangible feedback. Advances in material sciences and flexible display technologies are investigated to enable such reconfigurable physical objects. While existing solutions aim for making physical objects more controllable via the digital world, we propose an approach where holograms (virtual objects) in a mixed reality environment are augmented with physical variables such as shape, texture or temperature. As such, the support for mobility forms an important contribution of the  proposed solution since it enables users to freely move within and across environments. Furthermore, our augmented virtual objects can co-exist in a single environment with programmable matter and other actuated shape-changing solutions. The future potential of the proposed approach is illustrated in two usage scenarios and we hope that the presentation of our work in progress on a novel way to realise tangible holograms will foster some lively discussions in the CHI~community.
\end{abstract}

\keywords{\plainkeywords}


\section{Introduction and Related Work}
In their seminal work on Tangible Bits~\cite{Ishii1997}, Ishii and Ullmer presented a vision on bridging the gap between the digital and physical worlds via tangible user interfaces~(TUIs). Various subsequent projects, including the work on \mbox{Cartouche} by Ullmer~et~al.~\cite{Ullmer2010}, introduced new frameworks and conventions for building tangible user interfaces. However, in this first wave of research on tangible human-computer interaction, TUIs mainly acted as input controls for digital information and services while output was mainly provided via traditional computer screens.

\begin{marginfigure}
  \begin{minipage}{\marginparwidth}
  \vspace{-4cm}
	\includegraphics[width=0.9\marginparwidth]{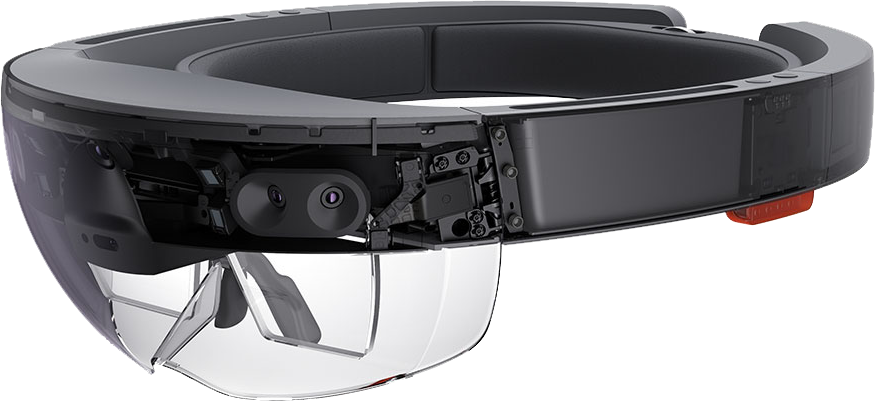}
  \caption{Microsoft HoloLens}
  \label{fig:hololens}
  \end{minipage}
\end{marginfigure}
 
The recent work on Radical Atoms by Ishii~et~al.~\cite{Ishii2012} plans to overcome the limitation that Tangible Bits are mainly used as handles for tangible input by moving towards transformable materials and shape-changing interfaces. A user can no longer just update the underlying digital model by interacting with a physical object, but changes in the underlying digital model are also communicated to the user by updating and changing the physical material (e.g.~shape). In their vision-driven design research, Ishii~et~al.~assume that future Radical Atoms will provide some form of sensing, actuation and communication at a molecular level. They introduce the hypothetical, reprogrammable and clay-like Perfect Red material which enables digital information to have its synchronised physical manifestation. Since the envisioned programmable matter is not currently available, a number of prototypes, simulating certain aspects of these envisioned materials have been realised, allowing for early exploration. For example, prototypes such as inFORM~\cite{Follmer2013} and TRANSFORM~\cite{Ishii2015} enable experimentation with shape-changing interfaces and dynamic physical affordances~\cite{Follmer2015}.

The digital model underlying any dynamic physical representation can of course not only represent simple objects but also complex data sets. In this case we are no longer talking about simple physical UI components but are rather dealing with data physicalisation where the underlying data can not only be explored visually but also via other physical properties. While the reprogrammable Perfect Red material proposed by Ishii~et~al.~is mainly about changing the shape of a physical artefact, Jansens~\cite{Jansen2015} proposes the use of other physical properties, called physical variables, such as smoothness, hardness or sponginess for providing additional feedback in data physicalisation. She further discusses some of the data physicalisation opportunities and challenges including the reconfigurability of physicalisations.

A drawback of many existing prototypes dealing with the simulation shape-changing materials (e.g.~inFORM and TRANSFORM) is their lack of mobility which means that they can only be used in situ at fixed places. Other solutions only work in virtual reality environments and again can only be used at fixed locations~\cite{Araujo2016}. On the other hand, we recently see the emergence of mobile devices such as the Microsoft HoloLens\footnote{https://www.microsoft.com/microsoft-hololens/} shown in Figure~\ref{fig:hololens}, enabling the development of realistic augmented reality environments. These solutions support the perfect embodiment of holograms (virtual objects) in physical space but when interacting with these environments it immediately becomes clear that the visual perception is not enough and something is missing since we cannot touch any embodied virtual object. As stated by Jansen:~\emph{\dquote{The perceived congruence concerns both visual as well as tactile perception. A perfect hologram will appear embodied but trying to touch it will destroy the illusion.}}\cite{Jansen2014}.


We present an alternative approach for simulating programmable matter based on the physical augmentation of virtual objects. Thereby we provide the missing physical features of holograms in mixed reality environments by proposing a wearable system that can be used to produce tangible holograms. In the remainder of this paper, we provide some details about the proposed wearable solution for tangible holograms, outline our ongoing work on a first prototype of the system and illustrate the potential of our solution based on two usage scenarios.


\section{Tangible Holograms}
As described in the previous section, there are currently a number of research prototypes for simulating programmable matter in order to investigate the future potential of shape-changing interfaces and interactions. Existing solutions start with the physical object and try to make the physical material and interfaces more configurable in order to enable dynamic physical affordances and support dynamic data physicalisation. We take an alternative approach and start with holograms that are perfectly embedded in physical environments. Rather than making physical objects more configurable, our challenge is therefore to add physical features to the already perfectly configurable digital holograms, without introducing too many limiting constraints in terms of mobility and other factors.

\begin{figure}[htb]
  \centering
  \includegraphics[width=0.85\columnwidth]{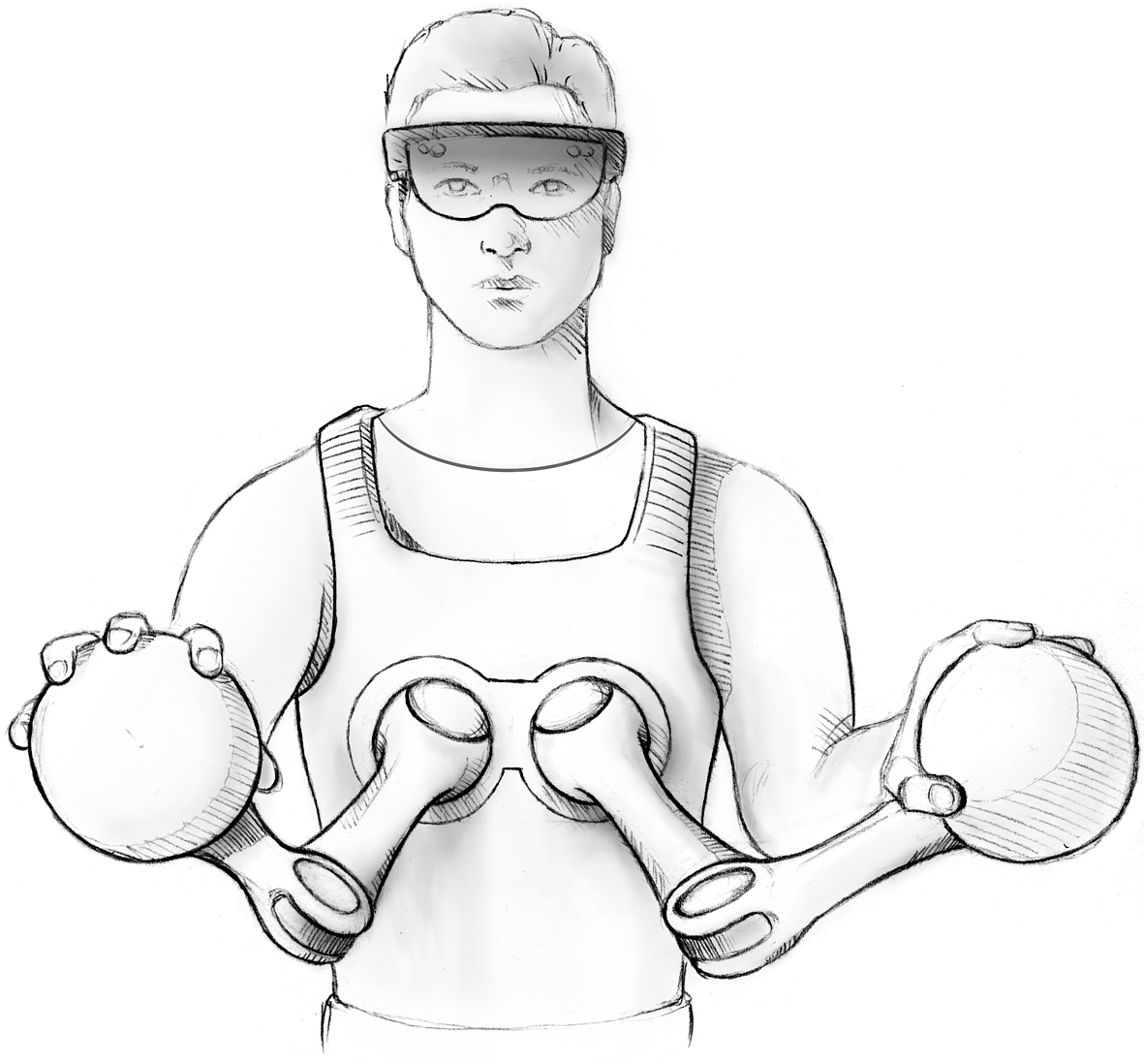}
	\caption{Tangible hologram system setup}
  \label{fig:setup}
\end{figure}

The setup of our proposed solution for mobile tangible holograms is shown in Figure~\ref{fig:setup}. First, the user has to wear some special glasses which will not only augment the user's view of the physical environment with the necessary digital holograms but also offer the possibility to track the spatial layout of the environment as well as any physical objects via depth and environmental cameras. The glasses can further track a user's hand in order to control any interaction with a hologram and to physically augment the hologram.

A second major component is a wearable vest or harness with two mounted robotic arms which are operated in front of a user's chest, similar to the body-attached robotic components described in other projects~\cite{Parietti2016,Leigh2016}. The end of each of the two robot arms carries a sphere which is used to capture the input from as user's hand or finger but also serves as an output device for different physical features. 

\begin{figure}[htb]
  \centering
  \includegraphics[width=\columnwidth]{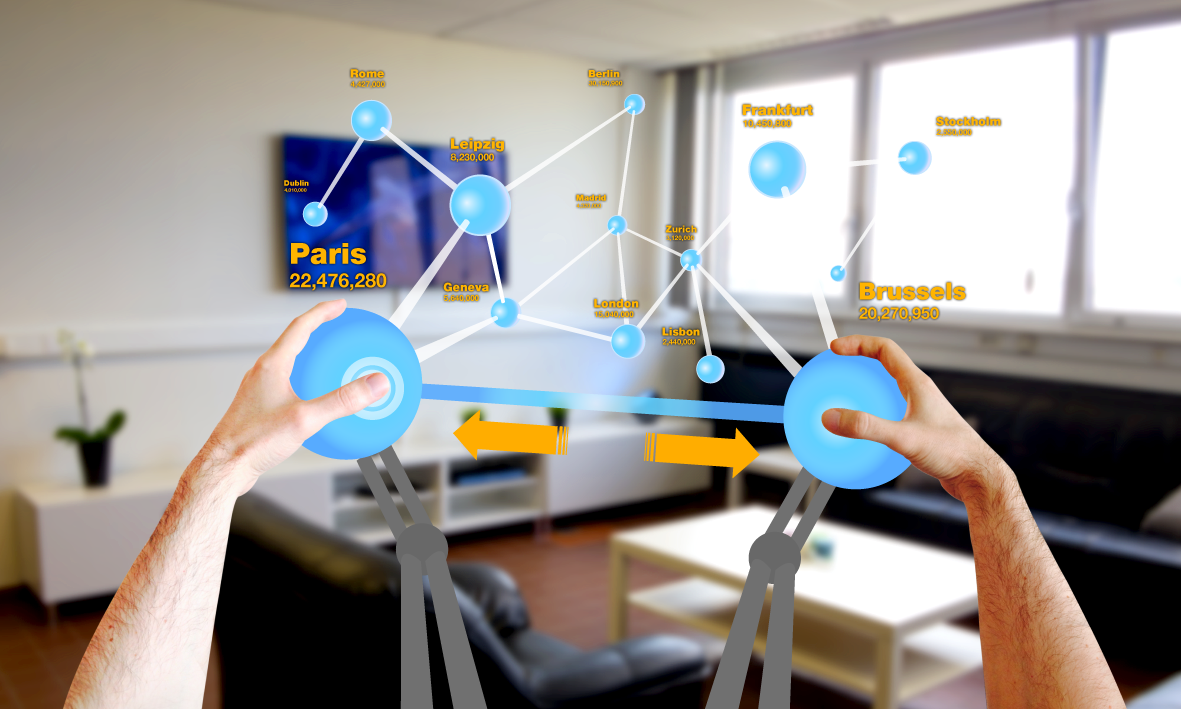}
  \caption{Interaction with a tangible graph hologram}
  \label{fig:mockup}
\end{figure}

As previously mentioned, the central idea is to physically augment holograms in a mixed reality environment. In order to do so, the spheres on the two robotic arms will be aligned with specific parts of the holograms in the environment based on the tracking of the cameras in the glasses. A first-person view mockup of a user interacting with a graph hologram is shown in Figure~\ref{fig:mockup}. The user's hands are positioned on the spheres and by moving the spheres the user can interact with the graph. Of course the user is free to move around in the room and can interact with any other parts of the graph since continuous tracking ensures that the spheres are re-aligned with other parts of the graph in real time. The spheres can not only be used as input devices but the robotic arms may also apply some directional force to the spheres in order to provide some additional computer-generated haptic force feedback based on the underlying digital model---as seen in other projects~\cite{Hurmuzlu1998}. In addition to the haptic feedback via the robot arms, the spheres can provide non-visual supplemental feedback about the underlying digital model or data via physical variables such as shape, texture or temperature as detailed later. Furthermore, other users wearing the same system might join the scene and collaboratively interact with the graph hologram. The interaction of one user may change the underlying digital model and thereby result in visual as well as physical feedback for the other users interacting with the same graph. Note that this represents a classical model-view-control~(MVC) design pattern with the peculiarity that views can no longer only be digital but also have a physical manifestation.

We see our solution as a modular platform for research on shape-changing interfaces, programmable matter and different physical variables. This platform can be customised with different interaction endpoints (spheres) depending on the application domain and the necessary feedback. Thereby, each sphere may be equipped with various sensors and actuators as realised in other tangible user interfaces~\cite{Poupyrev2007,Araujo2016}. While the spheres will act as simulators for various physical features of programmable matter, in the future they might even be built out of the Radical Atoms envisioned by Ishii~et~al.

\begin{marginfigure}
	\begin{minipage}{\marginparwidth}
	\vspace{1cm}
   \begin{subfigure}{.9\marginparwidth}
  \includegraphics[width=0.9\marginparwidth]{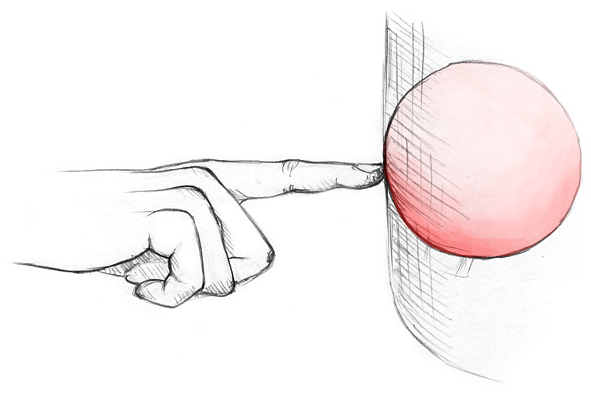}
 \subcaption{Pointing}
 \label{fig:pointing}
\end{subfigure} \\
\vspace{0.4cm}
\begin{subfigure}{.9\marginparwidth}
  \includegraphics[width=0.72\marginparwidth]{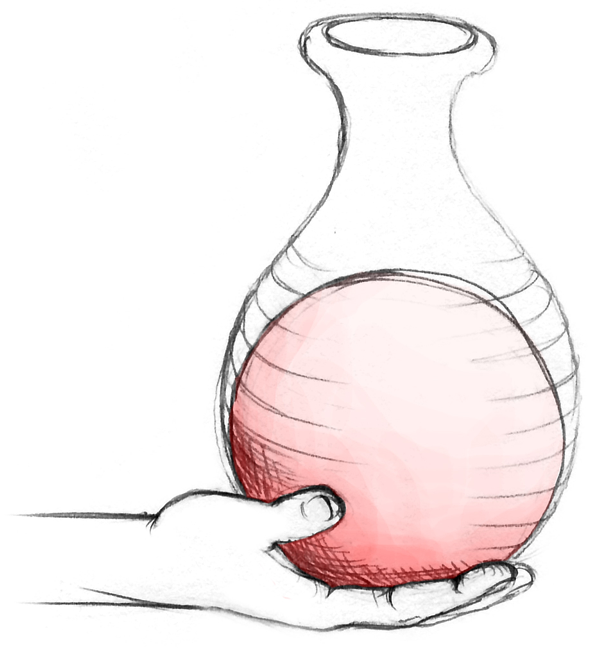}
\subcaption{Weighting}
\label{fig:weighting}
\end{subfigure} \\
\vspace{0.4cm}
\begin{subfigure}{.9\marginparwidth}
  \centering
  \includegraphics[width=0.9\marginparwidth]{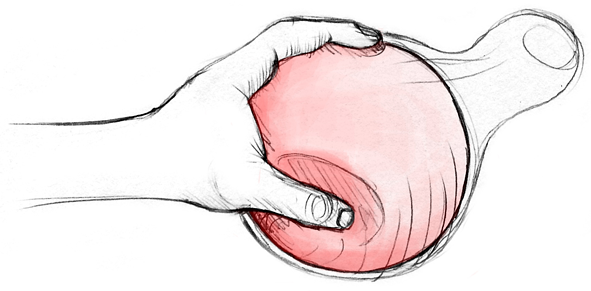}
\subcaption{Squeezing}
\label{fig:squeezing}
\end{subfigure}
\caption{Potential sphere interactions}
\label{fig:sphere}
  \end{minipage}
\end{marginfigure}

A number of different interactions can be supported by different spheres as shown in Figure~\ref{fig:sphere}. While the spheres have been introduced as tangible handles that can be used to interact with an underlying digital model and get physical feedback, Figure~\ref{fig:pointing} shows an alternative use where a sphere is used to simulate an arbitrarily shaped hologram. If a user moves their finger towards any part of a hologram, the tracking system in the glasses tracks the finger and quickly re-positions the sphere in order that its surface matches the surface of the point where the user's finger will touch the hologram. If we further ensure that the holographic projection occludes the shape of the sphere then the user experiences the illusion of a physical and tangible hologram. When the user moves their hand or finger, the sphere will continuously be repositioned to the next possible touch point and the resulting tangible hologram ensures that a touch can no longer destroy the illusion of the hologram. A similar solution for the simulation of tangible objects has been realised in the non-mobile VR Haptic Feedback Prototype\footnote{https://www.youtube.com/watch?v=iBWBPbonj-Q} for virtual reality environments. 

Another potential sphere interaction and interaction with a tangible hologram is shown in Figure~\ref{fig:weighting} where a user is trying to get an idea of the weight of a holographic vase. Again, the sphere is occluded by the holographic projection and the robot arm ensures that the sphere is producing the right level of pressure against the user's hand in order that they get a feeling about the vase's weight. Note that in addition the sphere might provide feedback about other physical features of the vase including its texture. 

Figure~\ref{fig:squeezing} shows a third possible interaction with one of the spheres where the user is squeezing the sphere and thereby producing some detailed input for the interaction with a tangible hologram. In addition, by squeezing a specific part they might get some feedback about the tangible hologram's physical variables such as elasticity or sponginess.

We are currently developing a first working prototype of the proposed system for tangible holograms. For the holographic glasses and the tracking of the environment as well as physical objects we are using a developer version of the Microsoft HoloLens discussed in the previous section. The realisation of the robotic arm prototype is based on Lego Mindstorms\footnote{https://www.lego.com/en-us/mindstorms} since this platform provides a flexible space within which to develop. A number of small serial link manipulators have been prototyped using this platform thus far, allowing us to begin evaluating several programming options. These include leJOS\footnote{http://www.lejos.org}, which allows for programming in Java, and a number of MATLAB toolboxes including the Robotics System Toolbox\footnote{https://www.mathworks.com/products/robotics.html}. Our final prototype will see a lightweight 3D~printed outer body replace most of the structural Lego elements and this structure will house the original Lego gear trains and servo motors. Note that this lightweight solution will add to the system's mobility. 
\section{Usage Scenarios}
In order to illustrate the potential of the described solution for tangible holograms we present a number of possible future usage scenarios. 

\begin{figure}[htb]
  \centering
  \includegraphics[width=0.9\columnwidth]{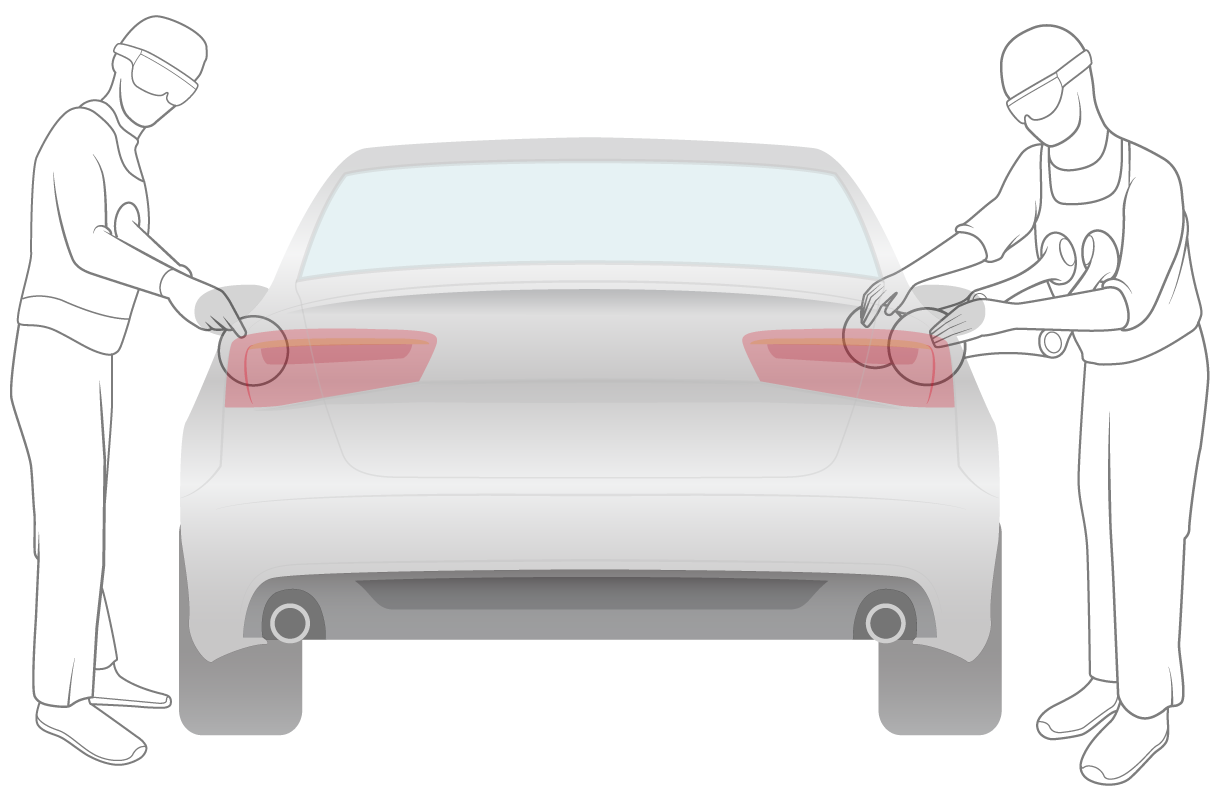}
  \caption{Collaborative work on a tangible car hologram}
  \label{fig:automotiveDesign}
\end{figure}

\textbf{Automotive Designer Scenario:} Two automotive designers are working on a 1/4 scale physical clay model of a car and wish to explore some refinements and design alternatives at full scale. Both designers are wearing our system and one makes a spatial mapping of the model using the head mounted unit, which is then presented to the user as a hologram---ready for manipulation. The first designer places the hologram in the centre of the room and then, from a short distance back, scales the model up to full size. This hologram is shared with the other designer and both begin to work on the model through natural, touch-based interactions as illustrated in Figure~\ref{fig:automotiveDesign}. The model can seem to take on the physical properties of clay, allowing it to be worked on and manipulated in the same manner as with the original physical medium but with the added benefit of having full control over properties such as plasticity.

After some time one designer decides to work on the model in another part of the studio and so they make a scaled-down copy of the hologram to bring with them. Both designers can now work on the model remotely and see each others manipulations as they happen, interacting, in effect, with two separate tangible holograms sharing the same underlying digital model.


\textbf{Interior Designer Scenario:} An interior designer wishes to discuss some design options with clients at an apartment in a block still under development. She and her two clients are wearing the system. In a given room they decide to explore some options and, accordingly, the interior designer performs a spatial mapping scan of the room, producing a mesh which is then overlayed with the room itself. This overlay may then be textured with any material (e.g.~a wallpaper pattern) and the resulting view is presented to both the designer and her clients. The designer then begins to realise the design options by scrolling through furnishings and placing items throughout the room as required. The designer and her clients can explore and experience the current design choices through the sense of touch as well as through vision. The clients can feel the texture of the seating, curtains and other furnishings. After the elements are initially positioned, the designer can make adjustments by directly touching and manipulating items.


\section{Conclusion and Future Work}
We have presented a new approach for simulating reconfigurable physical objects via tangible holograms. The combination of holograms, created by a Microsoft HoloLens, and a pair of wearable robotic arms with spherical controllers enables innovative ways of interacting with tangible holograms and experiencing their numerous physical variables (e.g.~shape, texture or temperature). While we are currently developing our own body-mounted robotic arm solution, in the future we might also investigate alternative mobile solutions for physical feedback, such as small drones that could carry our interaction spheres~\cite{Gomes2016}. We further plan to experiment with different sensors and actuators in our spheres for simulating different physical features. The proposed tangible hologram solution also differs from other systems for simulating programmable matter in the sense that we offer a self-contained and mobile solution that can be used anywhere, within as well as across environments. The automatic repositioning of the spheres based on a user's hand movements and the underlying digital model in combination with the superimposed projected holograms further enables the simulation of arbitrary physical shapes and features. While we have illustrated the potential of our approach in two multi-user usage scenarios, we see various opportunities how our tangible hologram system can be used in data physicalisation. The realisation of our tangible holograph prototype is ongoing work, but we hope that the presentation of the basic ideas of our novel mobile approach for the physical augmentation of virtual objects and the presented usage scenarios will lead to some constructive discussions among HCI researchers working on mixed reality, tangible interaction as well as data physicalisation.



\balance{} 

\bibliographystyle{SIGCHI-Reference-Format}
\bibliography{signer2017}


\begin{thebibliography}{00}


\ifx \showCODEN    \undefined \def \showCODEN     #1{\unskip}     \fi
\ifx \showDOI      \undefined \def \showDOI       #1{{\tt DOI:}\penalty0{#1}\ }
  \fi
\ifx \showISBNx    \undefined \def \showISBNx     #1{\unskip}     \fi
\ifx \showISBNxiii \undefined \def \showISBNxiii  #1{\unskip}     \fi
\ifx \showISSN     \undefined \def \showISSN      #1{\unskip}     \fi
\ifx \showLCCN     \undefined \def \showLCCN      #1{\unskip}     \fi
\ifx \shownote     \undefined \def \shownote      #1{#1}          \fi
\ifx \showarticletitle \undefined \def \showarticletitle #1{#1}   \fi
\ifx \showURL      \undefined \def \showURL       #1{#1}          \fi

\bibitem{Araujo2016}
{Bruno Araujo}, {Ricardo Jota}, {Varun Perumal}, {Jia~Xian Yao}, {Karan Singh},
  {and} {Daniel Wigdor}. 2016.
\newblock \showarticletitle{{Snake Charmer: Physically Enabling Virtual
  Objects}}. In {\em Proceedings of TEI 2016, International Conference on
  Tangible, Embedded and Embodied Interaction}. Eindhoven, The Netherlands,
  218--226.
\newblock
\showDOI{%
\url{http://dx.doi.org/10.1145/2839462.2839484}}


\bibitem{Follmer2015}
{Sean Follmer}. 2015.
\newblock {\em {Dynamic Physical Affordances for Shape-changing and Deformable
  User Interfaces}}.
\newblock Ph.D. Dissertation. Massachusetts Institute of Technology.
\newblock
\showURL{%
\url{http://hdl.handle.net/1721.1/97973}}


\bibitem{Follmer2013}
{Sean Follmer}, {Daniel Leithinger}, {Alex Olwal}, {Akimitsu Hogge}, {and}
  {Hiroshi Ishii}. 2013.
\newblock \showarticletitle{{inFORM: Dynamic Physical Affordances and
  Constraints Through Shape and Object Actuation}}. In {\em Proceedings of UIST
  2013, 26th Annual ACM Symposium on User Interface Software and Technology}.
  St. Andrews, Scotland, UK, 417--426.
\newblock
\showDOI{%
\url{http://dx.doi.org/10.1145/2501988.2502032}}


\bibitem{Gomes2016}
{Antonio Gomes}, {Calvin Rubens}, {Sean Braley}, {and} {Roel Vertegaal}. 2016.
\newblock \showarticletitle{{BitDrones: Towards Using 3D Nanocopter Displays as
  Interactive Self-Levitating Programmable Matter}}. In {\em Proceedings of CHI
  2016, ACM Conference on Human Factors in Computing Systems}. Santa Clara,
  USA, 770--780.
\newblock
\showDOI{%
\url{http://dx.doi.org/10.1145/2858036.2858519}}


\bibitem{Hurmuzlu1998}
{Yildirim Hurmuzlu}, {Anton Ephanov}, {and} {Dan Stoianovici}. 1998.
\newblock \showarticletitle{{Effect of a Pneumatically Driven Haptic Interface
  on the Perceptional Capabilities of Human Operators}}.
\newblock {\em Presence: Teleoperators and Virtual Environments\/} {7}, 3
  (1998), 290--307.
\newblock
\showDOI{%
\url{http://dx.doi.org/10.1162/105474698565721}}


\bibitem{Ishii2012}
{Hiroshi Ishii}, {D{\'a}vid Lakatos}, {Leonardo Bonanni}, {and} {Jean-Baptiste
  Labrune}. 2012.
\newblock \showarticletitle{{Radical Atoms: Beyond Tangible Bits, Toward
  Transformable Materials}}.
\newblock {\em interactions\/} {19}, 1 (January 2012), 38--51.
\newblock
\showDOI{%
\url{http://dx.doi.org/10.1145/2065327.2065337}}


\bibitem{Ishii2015}
{Hiroshi Ishii}, {Daniel Leithinger}, {Sean Follmer}, {Amit Zoran}, {Philipp
  Schoessler}, {and} {Jared Counts}. 2015.
\newblock \showarticletitle{{TRANSFORM: Embodiment of "Radical Atoms" at Milano
  Design Week}}. In {\em Proceedings of CHI 2015 EA, ACM Conference on Human
  Factors in Computing Systems}. Seoul, Republic of Korea, 687--694.
\newblock
\showDOI{%
\url{http://dx.doi.org/10.1145/2702613.2702969}}


\bibitem{Ishii1997}
{Hiroshi Ishii} {and} {Brygg Ullmer}. 1997.
\newblock \showarticletitle{{Tangible Bits: Towards Seamless Interfaces between
  People, Bits and Atoms}}. In {\em Proceedings of CHI 1997, ACM Conference on
  Human Factors in Computing Systems}. Atlanta, USA, 234--241.
\newblock
\showDOI{%
\url{http://dx.doi.org/10.1145/258549.258715}}


\bibitem{Jansen2014}
{Yvonne Jansen}. 2014.
\newblock {\em {Physical and Tangible Information Visualization}}.
\newblock Ph.D. Dissertation. Universit{\'e} Paris-Sud.
\newblock
\showURL{%
\url{https://tel.archives-ouvertes.fr/tel-00981521}}


\bibitem{Jansen2015}
{Yvonne Jansen}, {Pierre Dragicevic}, {Petra Isenberg}, {Jason Alexander},
  {Abhijit Karnik}, {Johan Kildal}, {Sriram Subramanian}, {and} {Kasper
  Hornb{\ae}k}. 2015.
\newblock \showarticletitle{{Opportunities and Challenges for Data
  Physicalization}}. In {\em Proceedings of CHI 2015, ACM Conference on Human
  Factors in Computing Systems}. Seoul, Republic of Korea, 3227--3236.
\newblock
\showDOI{%
\url{http://dx.doi.org/10.1145/2702123.2702180}}


\bibitem{Leigh2016}
{Sang-won Leigh} {and} {Pattie Maes}. 2016.
\newblock \showarticletitle{{Body Integrated Programmable Joints Interface}}.
  In {\em Proceedings of CHI 2016, ACM Conference on Human Factors in Computing
  Systems}. Santa Clara, USA, 6053--6057.
\newblock
\showDOI{%
\url{http://dx.doi.org/10.1145/2858036.2858538}}


\bibitem{Parietti2016}
{Federico Parietti} {and} {Harry Asada}. 2016.
\newblock \showarticletitle{{Supernumerary Robotic Limbs for Human Body
  Support}}.
\newblock {\em IEEE Transactions on Robotics\/} {32}, 2 (2016), 301--311.
\newblock
\showDOI{%
\url{http://dx.doi.org/10.1109/tro.2016.2520486}}


\bibitem{Poupyrev2007}
{Ivan Poupyrev}, {Tatsushi Nashida}, {and} {Makoto Okabe}. 2007.
\newblock \showarticletitle{{Actuation and Tangible User Interfaces: The
  Vaucanson Duck, Robots, and Shape Displays}}. In {\em Proceedings of TEI
  2007, First International Conference on Tangible and Embedded Interaction}.
  Baton Rouge, USA, 205--212.
\newblock
\showDOI{%
\url{http://dx.doi.org/10.1145/1226969.1227012}}


\bibitem{Ullmer2010}
{Brygg Ullmer}, {Zachary Dever}, {Rajesh Sankaran}, {Cornelius Toole}, {Chase
  Freeman}, {Brooke Cassady}, {Cole Wiley}, {Mohamed Diabi}, {Alvin Wallace},
  {Michael DeLatin}, {Blake Tregre}, {Kexi Liu}, {Srikanth Jandhyala}, {Robert
  Kooima}, {Chris Branton}, {and} {Rod Parker}. 2010.
\newblock \showarticletitle{{Cartouche: Conventions for Tangibles Bridging
  Diverse Interactive Systems}}. In {\em Proceedings of TEI 2010, Fourth
  International Conference on Tangible, Embedded, and Embodied Interaction}.
  Cambridge, USA, 93--100.
\newblock
\showDOI{%
\url{http://dx.doi.org/10.1145/1709886.1709904}}


\end{thebibliography}

\end{document}